\documentclass[11pt]{article}

\usepackage[margin=1in]{geometry}
\usepackage{amsmath,amssymb}
\usepackage{graphicx}
\usepackage{booktabs}
\usepackage{siunitx}
\usepackage{microtype}
\usepackage{xcolor}
\usepackage{hyperref}
\usepackage[nameinlink,noabbrev]{cleveref}
\usepackage[numbers,sort&compress]{natbib}

\usepackage{placeins}

\hypersetup{
  colorlinks=true,
  linkcolor=blue!45!black,
  citecolor=blue!45!black,
  urlcolor=blue!45!black
}

\newcommand{\notefigure}[3][]{%
  \IfFileExists{#2}{%
    \includegraphics[#1]{#2}%
  }{%
    \fbox{\parbox[c][0.25\textheight][c]{0.90\linewidth}{%
      \centering\small
      Figure file not yet present:\\[3pt]
      \texttt{\detokenize{#2}}\\[8pt]
      Upload the final PDF to the \texttt{figures/} directory.
    }}%
  }%
}

\newcommand{\CeBr}{CeBr$_3$}
\newcommand{\Nphot}{N_{\mathrm{phot}}}
\newcommand{\Fthree}{F_{385}}
\newcommand{\sigmalambda}{\sigma_{\lambda}}

\title{Spectral Discrimination of Deposited Gamma-Ray Energies in a Simulated CeBr\(_3\) Scintillator}

\author{
Andrey Elagin\thanks{Present address: Bolingbrook, IL 60440, USA; 
Email: \texttt{elagintech@gmail.com}}
}

\date{}

\begin{document}
\maketitle

\begin{abstract}
We show that wavelength measurements of individual detected optical photons may provide additional information about gamma-ray energy deposited in a \CeBr{} crystal when the detected-photon-count distributions overlap for nearby gamma-ray energies. Monoenergetic 662 and 629~keV gammas are used in a Geant4 simulation of a $25\times25\times20~\mathrm{mm^3}$ \CeBr{} crystal. Assuming a light yield of $6.0\times10^4$ photons/MeV, a wavelength-independent photon-detection efficiency of 30\%, and a wavelength resolution of $\sigma_{\lambda}=40$~nm, we find that the fraction of photons reconstructed above 385~nm gives an event-level separation of \(\sim 2\) standard deviations between the 662 and 629~keV event populations selected within the same \(\sim 1\%\)-wide detected-photon-count interval. No timing or reconstructed interaction-position information is used. The result demonstrates, within the present simulation model, that event-dependent optical spectra can retain energy information beyond an undifferentiated photon count.
\end{abstract}

\section{Introduction}
\label{sec:introduction}

The energy deposited in a scintillator is conventionally estimated from the total detected light, represented by an integrated charge or a detected-photon count.  Timing information can be used to infer depth of interaction and subsequently correct for depth-dependent light-transport effects, but the wavelengths of individual detected photons within each event are usually not used for energy reconstruction. A general method for including per-photon wavelength information in event-by-event energy estimation has been described in Ref.~\cite{ElaginPatent2026}. Here we present a specific \CeBr{} embodiment to show that after selecting events with approximately the same detected photon count, the detected spectral shape allows for discrimination between two nearby deposited gamma-ray energies.

Gammas of different energies can have different interaction-depth distributions, different numbers and spatial arrangements of interaction sites, and different spatial distributions of deposited energy. These differences can lead to different optical path-length distributions for photons produced by the radiation interaction before they reach the photodetector surface. Wavelength-dependent self-absorption, re-emission, reflection, and other optical transport effects can make the detected optical spectrum dependent on the gamma-ray energy, including for events with similar detected photon counts.

The overlap between the emission and absorption bands in \CeBr{} makes its detected spectrum sensitive to optical transport. At the same time, \CeBr{} has a high scintillation light yield, with values near \(6\times10^4\) photons/MeV reported for bare crystals \cite{Quarati2013}, making it attractive for energy measurements based on total detected light. Previous measurements of X-ray-excited emission spectra of \CeBr{} samples with thicknesses of 0.25, 2.5, and 25~mm showed a thickness-dependent spectral modification \cite{Quarati2013}. Self-absorption and re-emission in Ce$^{3+}$-based bromide scintillators have also been studied separately \cite{terWeele2014}. The wavelength dependence of these processes can make the detected spectral shape sensitive to differences in optical transport associated with different gamma-ray energies, providing additional information for event-by-event energy estimation.

We simulate monoenergetic 629 and 662~keV gamma rays in a
$25\times25\times20~\mathrm{mm^3}$ \CeBr{} crystal. A \CeBr{} crystal
of these dimensions has previously been studied experimentally near
662~keV \cite{Ulyanov2017}. The lower gamma-ray energy was chosen to be
approximately 5\% below 662~keV. In the present simulation, the main
peaks of the detected-photon-count distributions are separated, but the
distributions overlap substantially within an intermediate photon-count
range. The analysis below therefore compares full-energy-deposition events from the two energies
within narrow intervals of detected photon count, reducing the
discrimination available from total light yield and testing the additional
discrimination provided by the detected optical spectrum.

Optical spectral information has previously been used for several purposes in scintillator detectors. In PET, phoswich detectors use differences between scintillator layers to obtain depth-of-interaction information, conventionally through differences in scintillation decay time or pulse shape \cite{Seidel1999,Chang2017}. Wavelength-based approaches have also been developed for PET, including the use of color-sensitive SiPMs to distinguish stacked scintillators with different emission spectra \cite{Shimazoe2017} and wavelength discrimination in two-layer phoswich detectors using optical filters \cite{Ullah2020WLD,Ullah2020Optimization}. Wavelength sorting has also been used to separate Cherenkov and scintillation light \cite{Kaptanoglu2019,Kaptanoglu2020}, and multicolor scintillator stacks have been investigated for energy-resolved X-ray imaging \cite{Min2025}. New spectrally resolving single-photon detector concepts are under development \cite{Young2023,LeonardColorSensor}.

\section{\CeBr{} Detector Simulation}
\label{sec:simulation}

The simulation uses Geant4 \cite{Agostinelli2003} with optical-photon transport in a monolithic $25\times25\times20~\mathrm{mm^3}$ \CeBr{} crystal. One $25\times25~\mathrm{mm^2}$ face, corresponding to the $z = +10$~mm surface, is treated as the sensor surface. The other crystal surfaces are modeled with a wavelength-dependent PTFE-like diffuse wrapping, with approximately 94\% reflectivity over 360-430~nm, decreasing to 84-89\% at shorter wavelengths and increasing to about 96\% at longer wavelengths. 

Monoenergetic gamma rays are generated at $(0,0,-11~\mathrm{mm})$, 1~mm in front of the center of the $z = -10$~mm crystal face, and directed along $+z$. Ten thousand events are generated at each of 629 and 662~keV primary gammas.

The intrinsic scintillation yield of \CeBr{} is set to $6.0\times10^4$ photons/MeV.
Scintillation time constant of 17~ns is used\footnote{Timing information is propagated in the simulation but is not used in the analysis presented here.}. 
The Birks constant is set to $k_B=0.01~\mathrm{mm/MeV}$ as an approximate treatment of the known non-proportional scintillation response of \CeBr{}. Cherenkov photon production is also included in the simulation. All optical photons are transported through the wavelength-dependent Geant4 optical model, including geometric light-collection effects and transport losses.

The optical model is parameterized to reproduce the published thickness-dependent modification of the \CeBr{} emission spectrum rather than being constructed from a complete set of directly measured wavelength-dependent optical constants. The prompt emission spectrum is based on the X-ray-excited spectrum measured for a 0.25~mm \CeBr{} sample \cite{Quarati2013}, with a small correction toward shorter wavelengths to account for residual self-absorption in the thin sample. Self-absorbed photons are re-emitted with the same Ce$^{3+}$ spectral shape used for the prompt emission \cite{DorenbosPrivate2026}. The wavelength dependence of the self-absorption length is adjusted so that the simulated spectra approximately reproduce the published thickness-dependent spectral trend \cite{Quarati2013}. The resulting wavelength-dependent absorption-length parameterization~\cite{ElaginOpticalTables2026} should not be interpreted as a direct measurement of the \CeBr{} absorption coefficient. The overall treatment of self-absorption and re-emission is consistent with published studies of Ce$^{3+}$-activated bromide scintillators \cite{Quarati2013, terWeele2014}.

\begin{figure}[hbt!]
    \centering
    \includegraphics[width=0.89\textwidth]{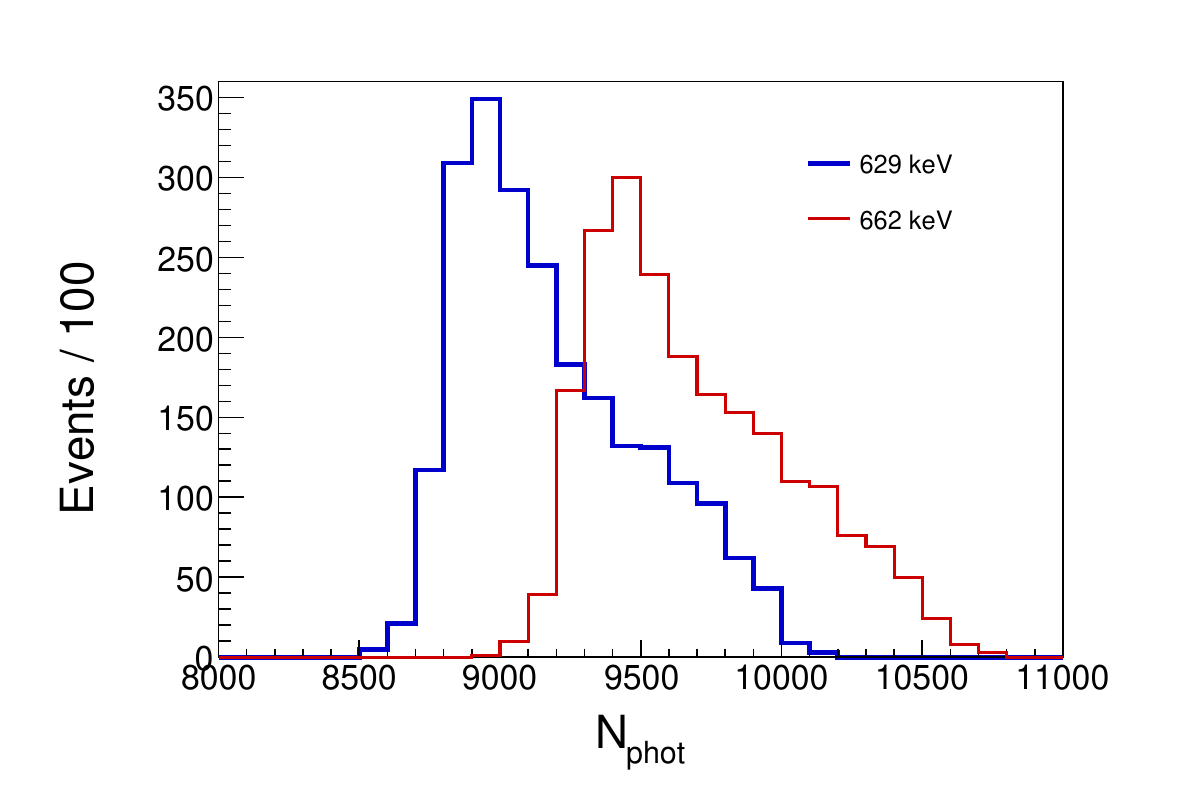}
    \caption{Number of detected photons for fully-deposited 629 and 662~keV gammas in a monolithic $25\times25\times20~\mathrm{mm^3}$ \CeBr{} crystal with the gammas entering from the $z = -10$~mm side. Of 10,000 simulated incident gammas at each energy, 2,268 events at 629~keV and 2,115 events at 662~keV are selected as full-energy-deposition events. Optical photons reaching the sensor surface at $z = +10$~mm are counted after wavelength-independent 30\% photon-detection-efficiency is applied. The remaining surfaces of the crystal are modeled with a wavelength-dependent PTFE-like diffuse wrapping.
    }
    \label{fig:hNphot}
\end{figure}

A wavelength-independent photon-detection efficiency of 30\%, representing a combined quantum-efficiency and collection-efficiency factor, is then applied independently to each photon reaching the $z=+10$~mm sensor surface. The number of photons accepted by this detector-response model in each event is denoted by $\Nphot$. These photons define the detected-photon population used in the analysis. The distribution of detected photons for fully-deposited events is shown in Fig.~\ref{fig:hNphot}.

\begin{figure}[hbt!]
    \centering
    \includegraphics[width=0.89\textwidth]{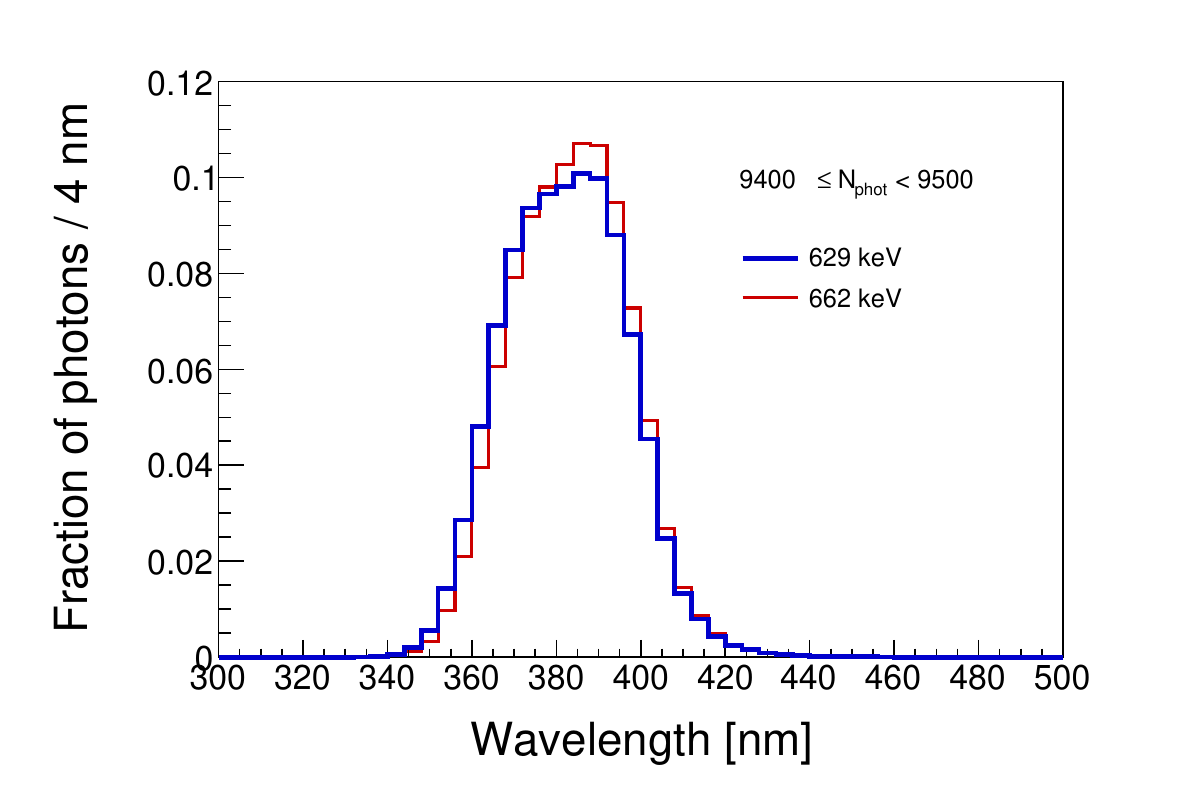}
    \caption{Normalized spectra of the detected photons for fully-deposited events from 629 and 662~keV gammas in the photon-count interval of $9400\leq\Nphot<9500$. In this interval, there are 132 events from 629~keV gammas and 300 events from 662~keV gammas. The spectra are normalized independently for the two gamma-ray energies. No per-photon wavelength smearing is applied in this figure. The plotted distributions, therefore, correspond to the true detected-photon wavelengths in the simulation.
    }
    \label{fig:hSpectrum}
\end{figure}

Figure~\ref{fig:hNphot} illustrates that events with nearby deposited energies may have overlapping photon counts. Simulation truth-level spectral differences for the detected photons in the counting interval near the peak of the 662~keV distribution, $9400\leq\Nphot<9500$, are shown in Fig.~\ref{fig:hSpectrum}. Although the two event samples are selected within the same narrow detected-photon-count interval, their normalized wavelength distributions are not identical. The 662~keV sample is shifted toward a larger fraction of photons at longer wavelengths, motivating the event-level spectral observable introduced below.

In the subsequent analysis, finite per-photon wavelength resolution is modeled by assigning
\begin{equation}
  \lambda_{\rm meas}=\lambda_{\rm true}+\delta\lambda,
  \qquad
  \delta\lambda\sim\mathcal{N}(0,\sigmalambda^2),
  \label{eq:smear}
\end{equation}
where $\lambda_{\rm true}$ is the true simulated wavelength of a detected
photon, $\lambda_{\rm meas}$ is the corresponding measured wavelength after
applying the detector-resolution model, and
$\mathcal{N}(0,\sigmalambda^2)$ denotes a Gaussian distribution with zero
mean and standard deviation $\sigmalambda$. A new value of $\delta\lambda$
is drawn independently for each photon, including photons from different
events. The baseline analysis uses $\sigmalambda=40$~nm.

\section{Data Analysis and Results}
\label{sec:results}

Full-energy deposition events are selected using the Monte Carlo deposited energy\footnote{This truth-level selection isolates the question of whether the detected spectral shape retains information about deposited energy when the total detected light is approximately the same, without introducing differences associated with partial energy escape.}. The selected events are then grouped into 100-photon-wide intervals of $\Nphot$. Around the representative interval $9400\leq\Nphot<9500$, this corresponds to approximately 1\% of a typical detected photon count. Conditioning on a narrow $\Nphot$ interval strongly reduces the discrimination available from total detected light, thus isolating spectral-based discrimination. 

The analysis in this section does not use photon arrival times for reconstructing interaction positions. Those observables can carry conventional information about light transport and depth of interaction, but excluding them here isolates the contribution from the event-specific spectral response.

For each event we define
\begin{equation}
  \Fthree =
  \frac{N(\lambda_{\rm meas}>385~\mathrm{nm})}
       {\Nphot}.
  \label{eq:f385}
\end{equation}
This intentionally simple observable uses only the fraction of detected photons measured above 385~nm and does not use the full measured wavelength distribution.

The separation between the 629 and 662~keV event populations in each photon-count bin is quantified by
\begin{equation}
  d' =
  \frac{\mu_{662}-\mu_{629}}
       {\sqrt{\left(\sigma_{629}^2+\sigma_{662}^2\right)/2}},
  \label{eq:dprime}
\end{equation}
where $\mu$ and $\sigma$ are the event-level mean and standard deviation of $\Fthree$. The quantity $d'$ measures the difference between the two population means relative to the widths of their $\Fthree$ distributions. We also calculate the empirical area under the receiver-operating-characteristic curve (AUC).  

\begin{figure}[hbt!]
    \centering
    \includegraphics[width=0.89\textwidth]{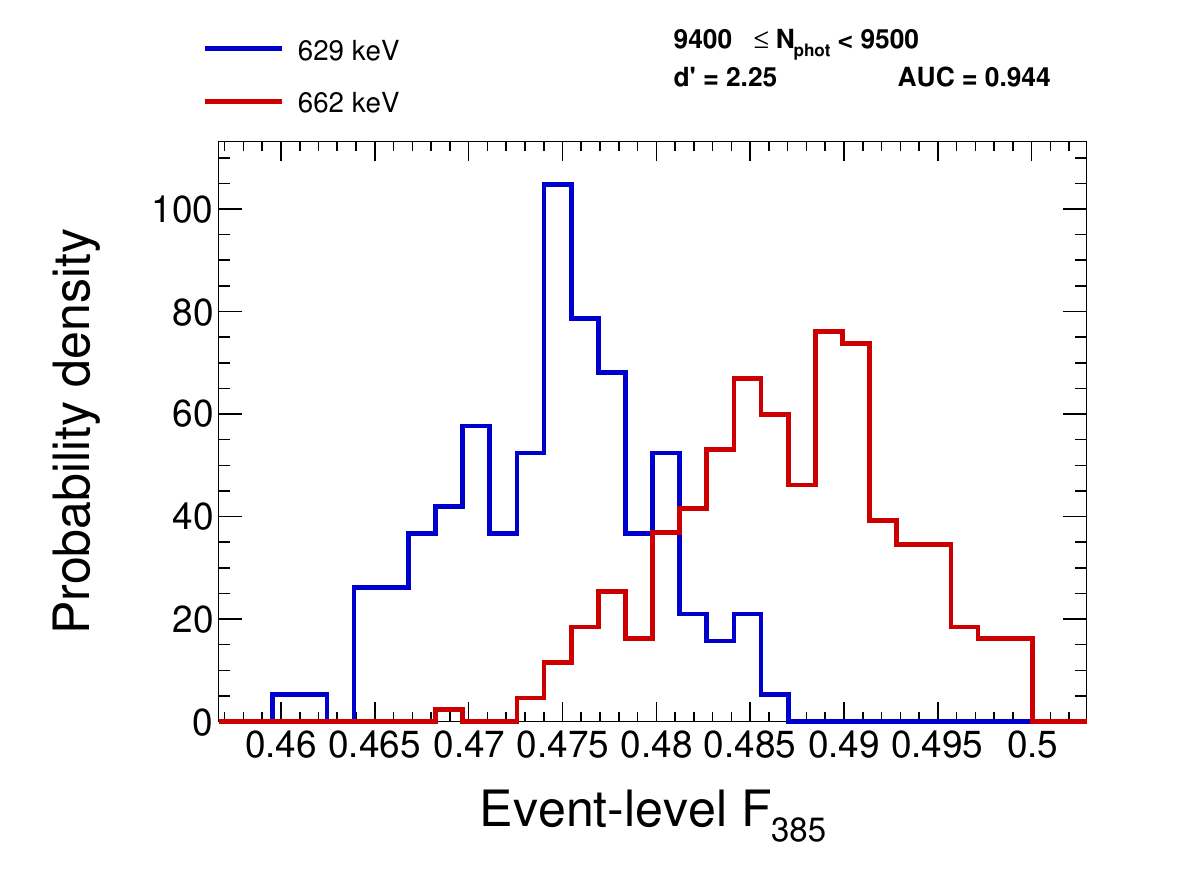}
    \caption{Event-level $\Fthree$ distributions for 629 and 662~keV full-energy events in the representative interval $9400\leq\Nphot<9500$, using independent per-photon wavelength smearing of $\sigmalambda=40$~nm.  Each event contributes a single $\Fthree$ value, and the histograms are normalized to probability density because the selected samples contain different numbers of events ($n=132$ at 629~keV and $n=300$ at 662~keV).  The population means are 0.4744 and 0.4871. The separation is $d'=2.25$, and the empirical AUC is 0.944. These quantities characterize event-population discrimination at approximately fixed detected-photon count.}
  \label{fig:f385dist}
\end{figure}

The event-level $\Fthree$ distributions in the representative
$9400\leq\Nphot<9500$ interval are shown in Fig.~\ref{fig:f385dist}.
This interval contains 132 selected 629~keV events and 300 selected 662~keV events, with each event contributing one value of $\Fthree$. With
$\sigmalambda=40$~nm, the mean $\Fthree$ values are 0.4744 and 0.4871,
respectively. Their difference, 0.0127, is small in absolute terms but is
resolved because each event contains approximately 9450 detected photons.
The corresponding population separation is $d'=2.25$, and the empirical
AUC is 0.944.

\begin{figure}[hbt!]
  \centering
  \includegraphics[width=0.89\textwidth]{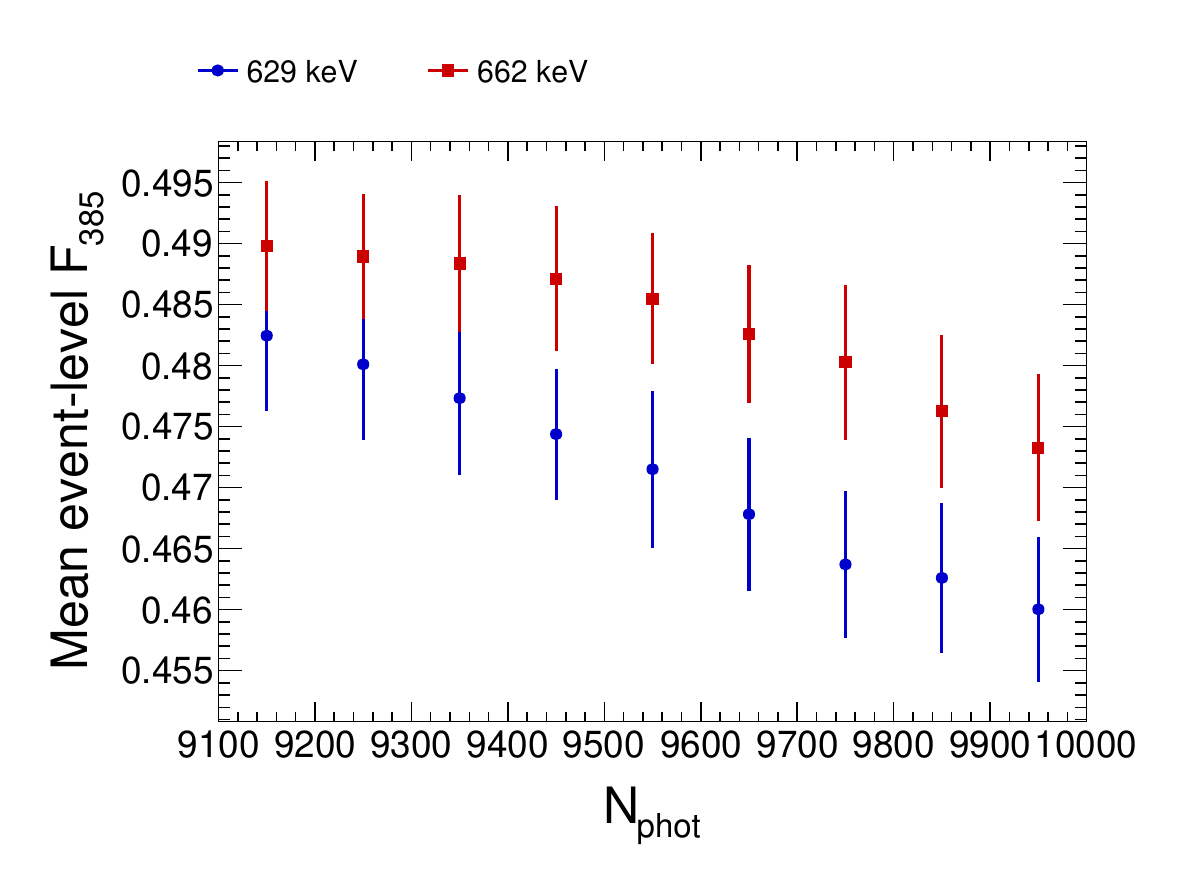}
  \caption{Mean event-level $\Fthree$ as a function of detected-photon count for full-energy 629 and 662~keV events, using $\sigmalambda=40$~nm. Events are grouped in 100-photon-wide bins. Vertical bars are $\pm1$ event-level standard deviation of $\Fthree$ in each population. The persistence of an offset between the two energy samples within the same narrow $\Nphot$ bins demonstrates that spectral information brings additional energy discrimination beyond conventional photon count.}
  \label{fig:f385mean}
\end{figure}

The conditional comparison is repeated over the range $9100\leq\Nphot<10000$ in Fig.~\ref{fig:f385mean}. The mean $\Fthree$ differs between the two energies throughout the analyzed range. The vertical bars show the event-level standard deviation of $\Fthree$.

\begin{figure}[hbt!]
  \centering
  \includegraphics[width=0.89\textwidth]{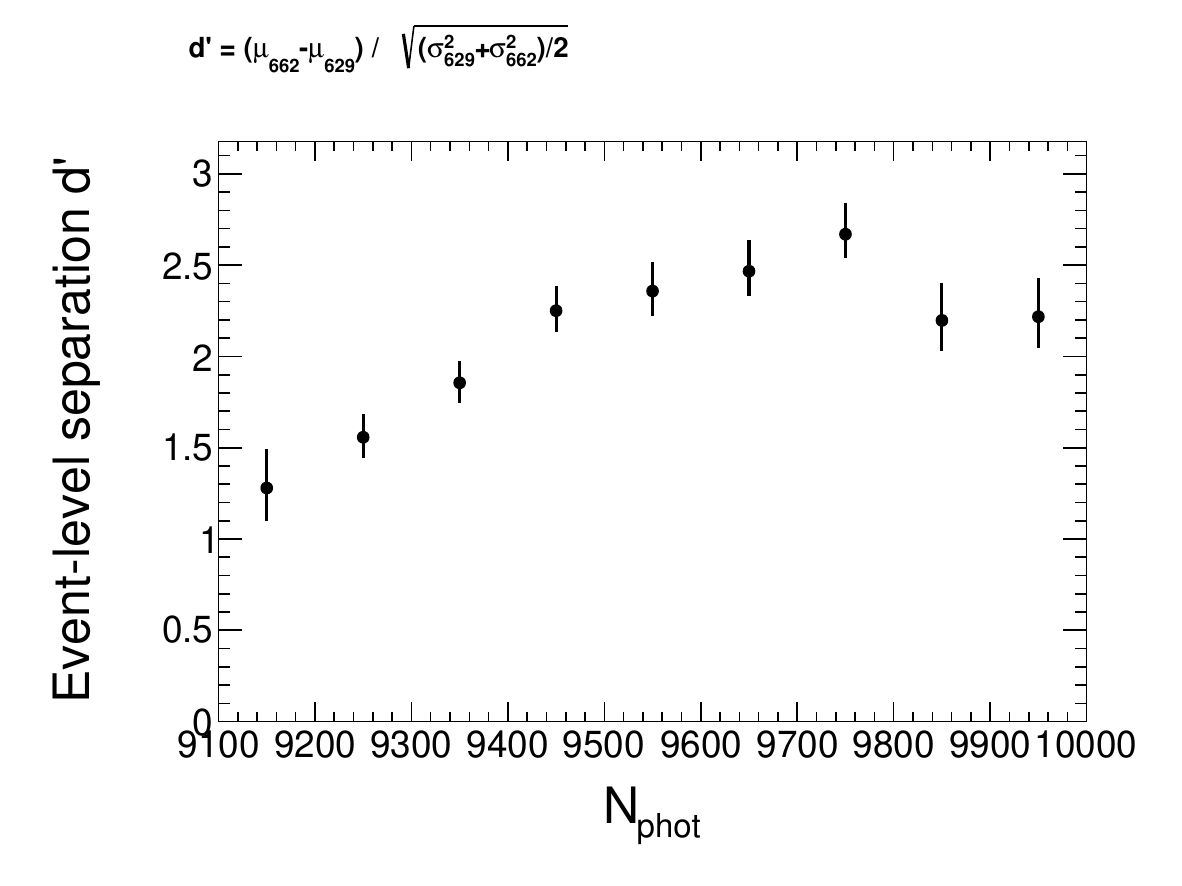}
  \caption{Event-population separation $d'$ between 629 and 662~keV as a function of detected-photon count for $\sigmalambda=40$~nm. The central points are calculated from the original event populations using Eq.~\eqref{eq:dprime}. Vertical bars show the central 68\% bootstrap interval obtained from event-level resampling within each photon-count bin.}
  \label{fig:dprime}
\end{figure}

Figure~\ref{fig:dprime} shows the event-population separation $d'$ as a function of detected-photon count. Uncertainties on $d'$ are obtained by independently resampling the 629 and 662~keV event populations with replacement within each $\Nphot$ bin. For the baseline $\sigmalambda=40$~nm model, $d'$ ranges from 1.28 to 2.67 over the nine analyzed photon-count bins. In the representative $9400\leq\Nphot<9500$ interval, $d'=2.25$, with a central 68\% bootstrap interval of 2.14-2.39.

\begin{figure}[hbt!]
  \centering
  \includegraphics[width=0.89\textwidth]{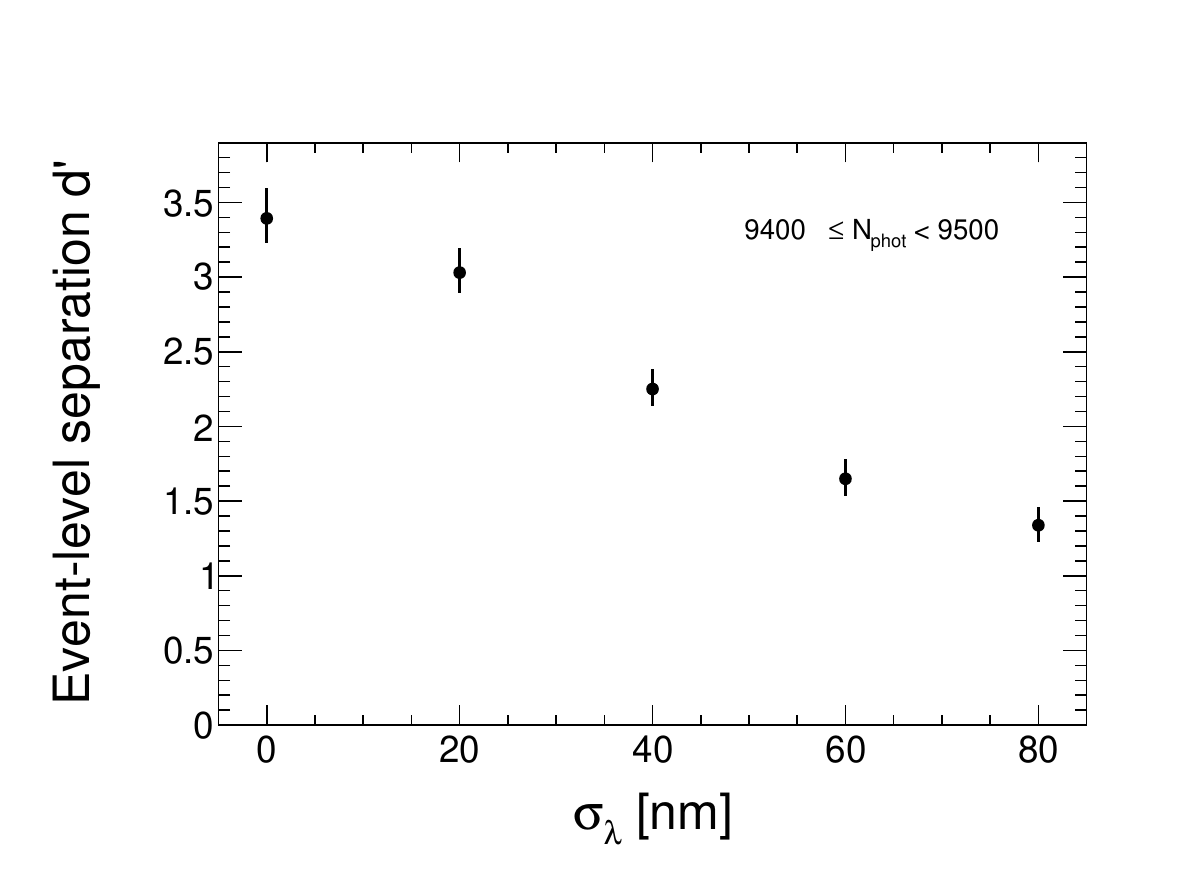}
  \caption{Event-population separation $d'$ as a function of the assumed
per-photon wavelength resolution for full-energy-deposition 629 and 662~keV events in the representative interval $9400\leq N_{\rm phot}<9500$. The event selection is held fixed while an independent Gaussian wavelength smearing with width $\sigma_\lambda$ is applied to each detected photon. Vertical bars show the central 68\% interval from event-level bootstrap resamples.}
  \label{fig:dprime_vs_sigma}
\end{figure}

The spectral separation depends on the assumed single-photon wavelength
resolution, as shown in Fig.~\ref{fig:dprime_vs_sigma}. Holding the event
selection and the $9400\leq N_{\rm phot}<9500$ photon-count interval fixed,
$d'$ decreases from 3.39 for perfect wavelength information to 3.03,
2.25, 1.65, and 1.34 for $\sigmalambda=20$, 40, 60, and 80~nm, respectively.

\FloatBarrier
\section{Outlook}
\label{sec:conclusions}

We demonstrated spectral differences of events corresponding to nearby energy depositions with similar photon count. The differences arise from wavelength-dependent optical transport in \CeBr{}, including self-absorption, re-emission, and reflection, which can produce different detected spectral shapes for events with nearby deposited energies. This spectral information may be used in combination with timing and position information to improve energy resolution in \CeBr{} and other materials where optical transport leads to sufficiently different spectra for events with nearby energy deposits.

The observable $\Fthree$ deliberately compresses every photon wavelength to a single binary decision at 385~nm. It therefore discards most of the spectral information available to a wavelength-resolving sensor. A likelihood constructed from the full per-photon wavelength distribution, a multi-bin spectral estimator, or a learned detector-specific response can use more of this information and should be evaluated in a complete reconstruction study. The present result establishes a simple reference point before introducing such machinery.

\section*{Disclosure of the Use of AI}
ChatGPT (OpenAI) was used as an editorial aid for language refinement, assistance with literature searches, and reference preparation. It was also used for code reviews and debugging purposes. It was not used to originate the scientific concept or perform the simulations described here. The author claims sole ownership of the scientific content presented herein and accepts full responsibility for the entire manuscript.

\clearpage

\bibliographystyle{unsrtnat}
\bibliography{references}

\end{document}